\documentclass[twocolumn,showpacs,preprintnumbers,amsmath,amssymb,epsf]{revtex4}
\usepackage{graphicx}
\usepackage{dcolumn}
\usepackage{bm}
\begin{document}
\title{Calculable lepton masses, seesaw relations and four neutrino mixings in a 3-3-1 model with extra U(1) symmetry}
\author{Nelson V. Cortez Jr.}
\affiliation{Rua Justino Boschetti 40, 02205-050 S\~ao Paulo, SP, Brazil}
\author{Mauro D. Tonasse}
\affiliation{Campus de Registro, Universidade Estadual Paulista, Rua Tamekishi Takano 5, 11900-000 Registro, SP, Brazil}
\date{\today}

\begin{abstract}
We propose a scheme in that the masses of the heavier leptons obey seesaw type relations. The light lepton masses, except the electron and the electron neutrino ones are generated by one loop level radiative corrections. We work in a version of the 3-3-1 electroweak model that predicts singlets (charged and neutral) of heavy leptons beyond the known ones. An extra U(1)$_\Omega$ symmetry is introduced in order to avoid the light leptons get masses at the tree level. The electron mass induces an explicit symmetry breaking at U(1)$_\Omega$. We discuss also the mixing matrix among four neutrinos. The new energy scale required is not higher than a few TeV. 
\end{abstract}

\pacs{14.60.Hi, 14.60.Pq, 14.60.St}
\maketitle

\section{\label{sec:level1}Introduction}
\label{sec1}

One of the main problems in particle physics consists in explaining the pattern of the fermion mass spectrum and mixing angles. In the standard model the charged fermion masses are generated through the Higgs coupling at the tree level. Thus, by tuning of the Yukawa coupling constants, it is possible to accommodate all the fermion masses and mixing angles in the theory as free parameters. However, fermion masses are spread in a large range. So we are compelled to assume Yukawa coupling from $\approx 10^{-6}$ for the electron until to $\approx 1$ for the top quark. It is an experimental fact that the charged fermion masses increase systematically from the first to the third generation. The standard model gives no explanation for this hierarchy and why the fermion masses, except the top quark one, are small relative to the electroweak scale ($v_W$ = 246 GeV). \par
Recent experiments on solar \cite{Aea02}, atmospheric \cite{Fea00} and terrestrial \cite{Eea03} neutrino oscillation considerably improved our understanding about neutrino masses and mixings. These laboratories experiments have been complemented by cosmological data, mainly by the satellite WMAP, that has put a limit on the sum of neutrino masses $\left[\sum m_i \leq \left(0.7 - 2.0\right) {\mbox{ eV }}\right]$ \cite{Bea03}. \par
In order to give a satisfactory explanation to these phenomena basically two types of models have been attempted: seesaw and radiative corrections. The seesaw mechanism is interesting because it establishes that one fermion mass is heavy at the expense of another to be light, leading to a route to understand the mass hierarchy problem \cite{MA98a,MA98b}. It is possible also that the seesaw schemes can be able to explain other very disparate scales in the framework of particle physics \cite{PE04}. However, the usual seesaw mechanism for small neutrino masses requires the introduction of a dimension-five operator and a high energy scale \cite{MA98a}. \par
On the other hand, it should be natural to interpret the light fermion masses as generated by higher orders of perturbative expansions \cite{MA90,MN89}. Both, seesaw and radiative correction schemes are possibilities that exist in extensions of the standard model. \par
Fermion mass generation by radiative corrections necessarily evolves the introduction of some set of symmetries. This is to avoid the light fermions from picking their masses from the tree level through their coupling to the Higgs boson multiplets with nonzero vacuum expectation values \cite{MN89}. Another good feature of the radiative generation mechanism is that the masses radiatively generated can be calculable as functions the heavy lepton ones \cite{GG73}. \par
In this work we examine these problems for the case of the leptons in the context of an electroweak model based on SU(3)$_L$$\otimes$U(1)$_N$ (3-3-1 for short) semi simple symmetry group \cite{PP92,FR92,FH93}. \par
There are in the literature several versions of the 3-3-1 model \cite{PP92,FR92,LO96}. Some of them can include mechanisms for neutrino mass generation \cite{MP01,MP02a,MP02b}. It is important to keep in mind that 3-3-1 model was not specifically proposed to solve the fermion mass problem. For the original motivations of the model see Refs. \cite{PP92,FR92}. \par
Here we show that in a variant of the model of the Refs. \cite{PP92,FR92} we can combine the good features from the seesaw mechanism and radiative corrections, which together are a very powerful tool, in order to obtain relations among the lepton masses. These relations are more general that the ones obtained by other models and, perhaps more importantly, the new energy scale is lower than the ones required by the other models. All the realizations of the usual seesaw models based on SU(2)$\otimes$U(1) symmetry require energies in the range $10^{13}$ GeV to $10^{16}$ GeV \cite{MA98a}. We will see that in our proposal we do not need a so high energy scale to implement the seesaw mechanism. \par
We discuss also the mixing angles for the neutrinos, but in our model the mixing occurs, in principle, among four neutrino states. However, the pattern of this mixing for the standard neutrino sector is of the LMA ({\it large mixing angle}) type. \par
Our work is organized as follow. In the Sec. \ref{sec2} we present the relevant aspects of the model. In the Sec. \ref{sec3} we describe the scheme for charged lepton mass generation, while in Sec. \ref{sec4} we present the model for neutrino masses generation and mixings. Finally in Sec. \ref{sec5} we have our conclusions and comments.

\section{\label{sec2}The model}

In the original version of the 3-3-1 model a scalar sextet was introduced, in addition to a set of three scalar triplets, in order to give the correct charged lepton mass pattern at the tree level \cite{FR92,FH93}. If the sextet is not present the $3 \times 3$ charged lepton mass matrix is antisymmetric, leading to one null and two degenerate mass eigenstates. However, was indicated a more simple solution to this problem, {\it i.e.}, we can eliminate the sextet of the scalar sector and to introduce an extra charged heavy lepton singlet in the leptonic sector \cite{DM93}. \par
Among all possible extensions of the standard model the class of the 3-3-1 models is one of most interesting. In this class of electroweak models the anomalies cancellation mechanism requires the number of fermion families to be a multiple of three. In addition, if we to take in account that the asymptotic freedom condition of QCD imposes that the number of generations is less than five, so the model predicts one unique three families group \cite{PP92,FR92}. Also, the Weinberg mixing angle is upper bounded in these models, {\it i. e.}, $\sin^2{\theta_W} < 1/4$. Therefore, the evolution of $\sin{\theta_W}$ to high values leads to an upper bound on the new mass scale for symmetry breaking of this model \cite{FR92,JJ97}. A feature of great phenomenological interest is that the model can manifest itself in a scale of several hundred of GeV or a few TeV. In addition, some of the 3-3-1 processes accessible to next generation of accelerators violates individual lepton numbers. For this reason, they are free of the standard model background \cite{MT02}. Therefore, 3-3-1 model is phenomenologically well motivated. \par
Here we work with the version of the model that includes a charged lepton singlet \cite{MP02b,DM93}. The lepton representation content in the interaction (primed) eigenstates is
\begin{subequations}
\begin{eqnarray}
\Psi_{aL} = \left(\begin{array}{ccc}\nu^\prime_a \cr \ell^\prime_a \cr {\ell^\prime}^C_a\end{array}\right)_L \sim \left({\bf 3}, 0\right), && \\
E^\prime_L, \quad E^\prime_R \sim \left({\bf 1}, -1\right), \quad N^\prime_R \sim \left({\bf 1}, 0\right), && 
\label{en}
\end{eqnarray}
\label{lep}\end{subequations}
where the subscript $a = e, \mu, \tau$ is a family index and $E^\prime_{L, R}$ are charged exotic lepton singlets. It should be noticed in Eqs. (\ref{en}) that, differently to the Ref. \cite{DM93}, we are including the neutral lepton singlet $N_R$ (see Ref. \cite{MP02b}). It is required by the recent experimental results on solar \cite{Aea02} and atmospheric \cite{Fea00} neutrino oscillations that indicate the presence of four neutrino states. \par
As in the original version, this model predicts two double charged gauge bosons, two single charged and one neutral in addition to the standard $W^\pm$ and $Z^0$ \cite{PP92,FR92}. The quark sector coincides with the one of the original model. \par
In order to give the fermion and gauge boson masses and account for the spontaneous symmetry breaking, the scalar triplets
\begin{equation}
\eta = \left(\begin{array}{ccc}\eta^0 \cr \eta_1^- \cr \eta_2^+\end{array}\right), \quad \rho = \left(\begin{array}{ccc}\rho^+ \cr \rho^0 \cr \rho^{++}\end{array}\right), \quad \chi = \left(\begin{array}{ccc}\chi^- \cr \chi^{--} \cr \chi^0\end{array}\right),
\label{tri}\end{equation}\noindent
transforming as $\left({\bf 3}, 0\right)$, $\left({\bf 3}, 1\right)$ and $\left({\bf 3}, -1\right)$, respectively, are introduced. We will see below that the $\eta$ triplet does not contribute to the masses of the leptons of the second and third generations. However, it is fundamental for the generation of the masses of the electron and the neutrinos. The neutral components of the scalars fields $\eta$, $\rho$ and $\chi$ develop the vacuum expectation values $\langle\eta^0\rangle = v$, $\langle\rho^0\rangle = u$ and $\langle\chi^0\rangle = w$ with $v^2 + u^2 = v_W^2$ and so, during the spontaneous symmetry breaking process these neutral fields are shifted as $\eta_0 \to v + \xi_\eta + i\zeta_\eta$, $\rho_0 \to u + \xi_\rho + i\zeta_\rho$ and $\chi_0 \to w + \xi_\chi + i\zeta_\chi$. \par
We wish only the third generation and the exotic lepton singlets $E$ and $N$ get masses at the tree level. Therefore, to avoid undesirable mass terms for the electron and the muon and its neutrinos at the tree level, we must eliminate their couplings through the $\eta$ scalar.
Thus, we define a charge whose operators are given by
\begin{equation}
\Omega_\varphi = \frac{\sqrt{3}}{3}\lambda_8 + X_\varphi ,
\label{carga}
\end{equation}\noindent
where $\varphi =$ $\Psi_L$, $E^\prime_{L, R}$, $N^\prime_R$, $\eta$, $\rho$, $\chi$ and $\lambda_8$ is the usual notation for the diagonal Gell-Mann matrix. In the following, we 
assign the values $X_\varphi = -1/3$ for $\varphi$ $=$ $\Psi$, $\eta$ and $\rho$, and $X_\varphi = 2/3$ for $\varphi$ $=$ $\chi$. In order to become easier to recognize all participating and forbidden interactions in the Lagrangians, we list all the $\Omega_\varphi$ charges for the fields in the Eqs. (\ref{lep}) and (\ref{tri}) in Table \ref{tab1}. 

\begin{table}[h]
\caption{\label{tab1} $\Omega_\varphi$ charges for the fields of the model. $\Omega_{\ell^\prime_{aR}}$ and $\Omega_{{\ell^{\prime C}_a}_L}$ $\left(a = e, \mu, \tau\right)$ are not connected by the charge conjugation operation.}
\begin{ruledtabular}
\begin{tabular}{c|ccccccccccc}
$\Omega_\varphi$&&&&&Fields&\\
\hline
0 & $\nu^\prime_{aL}$ & $\ell^\prime_{aL}$ & $\ell^\prime_{aR}$ & $E^\prime_L$ & $E^\prime_R$ & $N^\prime_R$ & $\eta^0$ & $\rho^0$ & $\chi^0$ & $\eta^+_1$ & $\rho^+$\\
$-1$& ${\ell^{\prime C}_a}_L$ & $\eta^+_2$ & $\rho^{++}$ & $\chi^+$ & $\chi^{++}$ & & & & & \\
\end{tabular}
\end{ruledtabular}
\end{table}
The scalar sector of this model is the same as the three triplet's model of the Refs. \cite{PT93a,TO96}.  The most general renormalizable Higgs potential satisfying the set of symmetry associated with the charge operator (\ref{carga}) is
\begin{widetext}
\begin{eqnarray}
V\left(\eta, \rho, \chi\right) & = & \mu_1^2\eta^\dagger\eta + \mu_2^2\rho^\dagger\rho + \mu_3^2\chi^\dagger\chi + \lambda_1\left(\eta^\dagger\eta\right)^2 + \lambda_2\left(\rho^\dagger\rho\right)^2 + \lambda_3\left(\chi^\dagger\chi\right)^2 + \cr
&& + \eta^\dagger\eta\left(\lambda_4\rho^\dagger\rho + \lambda_5\chi^\dagger\chi\right) + \lambda_6\left(\rho^\dagger\rho\right)\left(\chi^\dagger\chi\right) + \lambda_7\left(\rho^\dagger\eta\right)\left(\eta^\dagger\rho\right) + \cr
&& + \lambda_8\left(\chi^\dagger\eta\right)\left(\eta^\dagger\chi\right) + \lambda_9\left(\rho^\dagger\chi\right)\left(\chi^\dagger\rho\right) + \frac{\Lambda}{2}\left(\epsilon^{ijk}\eta_i\rho_j\chi_k + {\mbox{H. c.}}\right),
\label{pot}\end{eqnarray}
\end{widetext}
where $\mu_i$, $i = 1, 2, 3$ and $\Lambda < 0$ (from the positivity of a Higgs boson mass) are constants with dimension of mass and $\lambda_j$, $j = 1, \dots, 9$ are adimensional constants. 
It must be noticed that interaction $\left(\eta^\dagger\chi\right)\left(\eta^\dagger\rho\right)$, allowed by the gauge symmetry, is forbidden in the potential (\ref{pot}) by $\Omega_\varphi$. The scalar potential (\ref{pot}) respects also $L + B$ symmetry, where $L$ and $B$ are the lepton and the baryon numbers \cite{TO96,PT93}. Therefore, the eigenstate of the physical double charged Higgs boson $H^{++}$ is defined by
\begin{subequations}
\begin{equation}
\left(\begin{array}{cc}\rho^{++} \\ \chi^{++}\end{array}\right) = \frac{1}{\sqrt{u^2 + w^2}}\left(\begin{array}{cc}-u & w \\ w & u \end{array}\right)\left(\begin{array}{cc}G^{++} \\ H {++}\end{array}\right), 
\label{eigh}
\end{equation}\noindent 
and the single charged $H^+_1$ and $H^+_2$ are 
\begin{eqnarray} \left(\begin{array}{cc}\eta^+_1 \\ \rho^+\end{array}\right) & = & \frac{1}{v_W}\left(\begin{array}{cc}-v & u \\ u & v \end{array}\right)\left(\begin{array}{cc}G_1^+ \\ H_1^+\end{array}\right), \label{eigh1+}\\ \left(\begin{array}{cc}\eta^+_2 \\ \chi^+\end{array}\right) & = & \frac{1}{\sqrt{v^2 + w^2}}\left(\begin{array}{cc}-v & w \\ w & v \end{array}\right)\left(\begin{array}{cc}G_2^+ \\ H_2^+\end{array}\right),
\label{eigh2+}
\end{eqnarray}
where $G^{++}$, $G_1^+$ and $G_2^+$ are charged massless Goldstone bosons that are {\it eaten} by $U^{++}$, $W^+$ and $V^+$ charged gauge bosons, respectively, through the Higgs mechanism. \par
In the neutral sector we have
\begin{eqnarray}
\left(\begin{array}{cc} \xi_\eta \\ \xi_\rho \end{array}\right) & \approx & \frac{1}{v_W}\left(\begin{array}{cc} v & u \\ u & -v \end{array}\right)\left(\begin{array}{cc}H_1^0 \\ H_2^0\end{array}\right), \label{eighh}\\ \xi_\chi & \approx & H_3^0, \quad \zeta_\chi \approx h^0.
\label{eigh0}
\end{eqnarray}
\end{subequations}
The other two neutral states are pure Goldstone that are {\it eaten} by the $Z^0$ and $Z^{\prime0}$ neutral gauge bosons \cite{TO96}. The pattern of symmetry breaking is SU(3)$_L$\-$\otimes$\-U(1)$_N$\-$\stackrel{\langle\chi\rangle}{\longmapsto}\- $SU(2)$_L$\-$\otimes$\-U(1)$_Y$\-$\stackrel{\langle\eta, \rho\rangle}{\longmapsto}$\-U(1)$_{{\rm em}}$. Therefore, since the vacuum expectation value $w$ governs the masses of the new fields of the model it is natural to assume $w \gg v, u$. The eigenstates (\ref{eighh}) and (\ref{eigh0}) are valid in this approximation. \par
The zeroth order Yukawa Lagrangian for the charged leptons that respects the symmetries of the model, including the conservation of the charge give in (\ref{carga}), is
\begin{widetext}
\begin{equation}
-{\cal L}^{\left(0\right)}_\ell = \sum_{a = e, \mu, \tau}\left(f_a\overline{\Psi_{aL}}E^\prime_R\rho + f_a^\prime\overline{E^\prime_L}\chi^{\tt T}{\Psi^C}_{aR}\right) + M\overline{E^\prime_L}E^\prime_R + {\mbox{H. c.}}
\label{yuk}
\end{equation}
\end{widetext}
Here $M$ is a constant with dimension of mass and $f_a$ and $f_a^\prime$ are the arbitrary adimensional Yukawa coupling strengths. It should be noticed that, as occur in some processes in weak interactions, the charge conjugation operation is not respected in the Lagrangian (\ref{yuk}). Concerning the $\Omega_\varphi$ charge, looking to the term $\overline{E^\prime_L}\ell^\prime_{aR}\chi^0$ and to assure $\Omega_\varphi$ conservation we have assumed $\Omega_{\ell^\prime_{aR}} = 0$ (see Table \ref{tab1}). However, the $C$ operator, defined by ${\psi_L}^C = C\overline{\psi_R}^T$ requires $\Omega_{\ell^\prime_{aR}} = 1$, since $\Omega_{{{\ell^\prime_a}^C}_L} = -1$. \par
Similarly, for the neutrino sector we can write the zeroth order Lagrangian as
\begin{equation}
{\cal L}^{\left(0\right)}_\nu = \sum_{a = e, \mu, \tau}h_a\overline{\psi_{aL}}N^\prime_R\eta + m\overline{{N_R^\prime}^C}N^\prime_R + {\mbox{H. c.}},
\label{lnu}
\end{equation}
where $m$ is a scale of mass and $h_a$ are arbitrary Yukawa couplings. It must be noticed that in the leptonic sector the $\Omega_\varphi$ charge conservation forbids all the couplings {\it via} $\eta$ scalar, {\it i. e.} $\epsilon^{ijk}\overline{\left(\Psi_{a_Li}\right)^C}F_{ab}\Psi_{bLj}\eta_k$, for the anti-symmetric Yukawa couplings $F_{ab}$ with $i, j, k = 1, 2, 3$ \cite{MP02b}. \par
We have checked that the set of symmetries introduced by the charge operator (\ref{carga}) can be extended to the quark sector of the Yukawa Lagrangian.

\section{\label{sec3}Charged lepton masses}

At the tree level the charged lepton mass matrix coming from the Lagrangian (\ref{yuk}) is
\begin{equation}
{\cal M}_0^{(\ell)} = \left(\begin{array}{cccc}0 & 0 & 0 & f_eu \\ 0 & 0 & 0 & f_\mu u \\ 0 & 0 & 0 & f_\tau u \\ f^\prime_ew & f^\prime_\mu w & f^\prime_\tau w & M\end{array}\right),
\label{mat0}
\end{equation}\noindent
in the $\left(e^\prime, \mu^\prime, \tau^\prime, E^\prime\right)_{L,R}$ basis. \par
The lepton mass matrix (\ref{mat0}) gives the seesaw mass relation
\begin{equation}
m_Em_\tau = -\left(f_ef_e^\prime + f_\mu f^\prime_\mu + f_\tau f^\prime_\tau\right)uw, \label{tau}\end{equation}\noindent
between the masses $m_\tau$, of the tau lepton, and $m_E$ for the exotic one, with
\begin{equation}
2m_\tau = M - \sqrt{M^2 + 4uw\left(f_ef_e^\prime + f_\mu f_\mu^\prime + f_\tau f_\tau^\prime\right)}.
\label{etau}
\end{equation}\noindent
Actually, $m_\tau$ in Eq. (\ref{etau}) can be negative, but the signal of the lepton mass is meaningless. In fact we always can exchange the signal of a lepton mass by a $\gamma_5$ transformation on the Dirac spinor $\left(\psi \to \gamma_5\psi\right)$. Therefore, hereafter we will not lead in account the sign of the lepton masses in our discussion. \par
A seesaw relation between the tau and the heavy lepton mass similar to that in Eq. (\ref{tau}) is given in Ref. \cite{MP02b}, but in that model more Yukawa couplings are required varying in a larger range (see our numerical example below). In our scheme the extra U(1)$_\Omega$ symmetry, whose charge is given in (\ref{carga}), avoid coupling of the leptons through the $\eta$ scalar triplet and governs also the suppression of some other parameters. The muon and electron masses vanish at tree level, but the muon one can be induced at the first order of perturbative expansion, while the electron one can be responsible by a tiny break of the U(1)$_\Omega$ symmetry. In order to obtain higher order finite mass terms for the charged leptons which effectively contribute to mass matrix (\ref{mat0}) we diagonalize it by making the transformations
\begin{widetext}
\begin{subequations}
\begin{eqnarray}
e^\prime_L & = & \frac{1}{f}\left(\frac{f_ef_\mu}{F}e_L - f_\tau\mu_L\right) - \frac{f_e}{F}\left(s_{\theta_\ell}\tau_L - c_{\theta_\ell}E_L\right), \\
\mu^\prime_L & = & \frac{1}{F}\left[-fe_L + f_\mu\left(-s_{\theta_\ell}\tau_L + c_{\theta_\ell}E_L\right)\right], \\
\tau^\prime_L & = & \frac{1}{f}\left(\frac{f_\mu f_\tau}{F}e_L + f_e\mu_L\right) - \frac{f_\tau}{F}\left(s_{\theta_\ell}\tau_L - c_{\theta_\ell}E_L\right), \\
E^\prime_L & = & -c_{\theta_\ell}\tau_L - s_{\theta_\ell}E_L,
\end{eqnarray}
\label{left}
\end{subequations}
\end{widetext}
in the left-handed sector, where we are combining the parameters as $f^2 = f_e^2 + f_\tau^2$, $F^2 = f^2 + f_\mu^2$, $c_{\theta_\ell} = \cos{\theta_\ell}$ and $s_{\theta_\ell} = \sin{\theta_\ell}$. In the right-handed sector we define $f^{\prime2} = f_e^{\prime2} + f_\mu^{\prime2}$, $F^{\prime2} = f^{\prime2} + f_\tau^{\prime2}$, $c_{\phi_\ell} = \cos{\phi_\ell}$, and $s_{\phi_\ell} = \sin{\phi_\ell}$ and the relations between the physical and the symmetry eigenstates are given by
\begin{widetext}
\begin{subequations}
\begin{eqnarray}
e^\prime_R & = & -\frac{1}{f^\prime}\left(\frac{f^\prime_ef^\prime_\tau}{F^\prime}e_R + f^\prime_\mu\mu_R\right) - \frac{f^\prime_e}{F^\prime}\left(s_{\phi_\ell}\tau_R - c_{\phi_\ell}E_R\right), \\
\mu^\prime_R & = & -\frac{1}{f^\prime}\left(-\frac{f^\prime_ef^\prime_\tau}{F^\prime}e_R + f^\prime_\mu\mu_R\right) - \frac{f^\prime_\mu}{F^\prime}\left(s_{\phi_\ell}\tau_R - c_{\phi_\ell}E_R\right), \\
\tau^\prime_R & = & \frac{1}{F^\prime}\left[f^\prime e_R + f^\prime_\tau\left(-s_{\phi_\ell}\tau_R + c_{\phi_\ell} E_R\right)\right], \\
E^\prime_R & = & -c_{\phi_\ell}\tau_R - s_{\phi_\ell} E_R
\end{eqnarray}
\label{right}
\end{subequations}
with $\tan{\theta_\ell} = 2FMu/\kappa$, $\tan{\phi_\ell} = \kappa/\left(2F^\prime Mw\right)$ and 
\begin{equation}
\kappa = F^{\prime2}w^2 - F^2u^2 + M^2 - \sqrt{\left[M^2 + \left(Fu - F^\prime w\right)^2\right]\left[M^2 + \left(Fu + F^\prime w\right)^2\right]}.
\end{equation}
With the eigenstates in Eqs. (\ref{eigh}), (\ref{left}) and (\ref{right}) we construct the first order Yukawa Lagrangian
\begin{eqnarray}
{\cal L}^{\left(1\right)}_+ & = & \frac{1}{\sqrt{u^2 + w^2}}\left[Fw\left(s_{\theta_\ell}\overline{{\tau^C}_L} + c_{\theta_\ell}\overline{{E^C}_L}\right)\left(c_{\phi_\ell}\tau_R + s_{\phi_\ell}E_R\right)H^{++} + \right. \cr
&& \left. + F^\prime u\left(c_{\theta_\ell}\overline{\tau_L} + s_{\theta_\ell}\overline{E_L}\right)\left(-s_{\phi_\ell}{\tau^C}_R + c_{\phi_\ell}{E^C}_R\right)H^{--}\right] + {\mbox{H. c.}},
\end{eqnarray}
for the charged leptons, which contributes to the entries $\left(3,3\right)$, $\left(3,4\right)$, $\left(4,3\right)$ and $\left(4,4\right)$ of the tree level mass matrix (\ref{mat0}) (see Fig. 1). Therefore, we have the first order mass term given by
\begin{eqnarray}
\sigma & = & \frac{\Lambda FF^\prime u^2}{32\pi^2w}s_{\theta_\ell}s_{\phi_\ell}\left\{m_\tau c_{\theta_\ell}c_{\phi_\ell}\frac{m_H^2 - m_\tau^2\left[1 + 2\ln\left(m_H/m_\tau\right)\right]}{\left(m_H^2 - m_\tau^2\right)^2} 
 -m_Es_{\theta_\ell}s_{\phi_\ell}\frac{m_H^2 - m_E^2\left[1 + 2\ln\left(m_H/m_E\right)\right]}{\left(m_H^2 - m_E^2\right)^2}\right\}.
\label{sigma}\end{eqnarray}
\end{widetext}
In the construction of the mass terms in Eq. (\ref{sigma}) we have considered only the greater contribution of the trilinear Higgs coupling in Fig. 1, {\it i. e.}, $i\Lambda uw/\left(u^2 + w^2\right)$. The charged lepton mass matrix with the first order contribution is
\begin{equation}
{\cal M}_1^{(\ell)} = \left(\begin{array}{cccc}0 & 0 & 0 & f_eu \\ 0 & 0 & 0 & f_\mu u \\ 0 & 0 & \sigma & f_\tau u \\ f^\prime_ew & f^\prime_\mu w & f^\prime_\tau w & M\end{array}\right),
\label{mat1}
\end{equation}\noindent
where we are neglecting corrections to the zeroth order terms $f_\tau u$, $f_\tau^\prime w$ and $M$. Now, we diagonalize the mass matrix (\ref{mat1}) taking into account that $m_\mu \ll m_\tau, m_E$. We obtain
\begin{equation} m_\mu \approx -\frac{\left(f_ef^\prime_e + f_\mu f^\prime_\mu\right)uw\sigma}{m_\tau m_E + \left(m_\tau + m_E\right)\sigma},
\label{massmu}
\end{equation}
with $m_E$ and $m_\tau$ given in Eqs. (\ref{tau}) and (\ref{etau}). \par
In order to show that the scheme is able to give the correct charged lepton mass pattern let us take a numerical example. Let $f_a = f^\prime_a = 0.0660i$ $\left(a = e, \mu, \tau\right)$, $u = 245$ GeV, $w = 1000$ GeV, $m_E = 1800$ GeV and $m_H = 80$ GeV. With these numerical values we obtain $\Lambda = -860.260$ GeV, $M = 1801.78$ GeV, $m_\tau = 1.7778$ GeV and $m_\mu = 0.106$ GeV [see Eqs. (\ref{tau}), (\ref{etau}) and (\ref{massmu})]. \par
However, the electron mass, that is still zero at this level, does not rise by higher order of perturbative expansion in this scheme. But it can induce a tiny explicit breaking of the U(1)$_\Omega$ symmetry through couplings
\begin{equation}
{\cal L}^{\left(0\right)}_{\eta Y} = \frac{g_{e\mu}}{2}\left(\overline{e^\prime_R}\mu^\prime_L - \overline{\mu^\prime_R}e^\prime_L\right)\eta^0
\end{equation}
of the charged leptons {\it via} the $\eta$ triplet. Therefore, if $g_{e\mu} = 5.29 \times 10^{-4}$, we replace the mass matrix (\ref{mat1}), with the numerical values given in the previous paragraph, by
\begin{equation}
{\cal M}_2^{(\ell)} \approx \left(\begin{array}{cccc} 0 & 0.012 & 0 & 16.17i \\ -0.012 & 0 & 0 & 16.17i \\ 0 & 0 & 0.17 & 16.17i \\ 65.98i & 65.98i & 65.98i & 1801.78\end{array}\right).
\label{mat2}
\end{equation}
This mass matrix gives $m_e = 0.51$ MeV with small modifications in the previous values of $m_N$, $m_\tau$ and $m_\mu$, which are in good agreement with the Particle Data Group values \cite{Hea02}. \par
The charged lepton mixing matrices, in the $\left(e, \mu, \tau, E\right)$ basis, are
\begin{figure}
\includegraphics{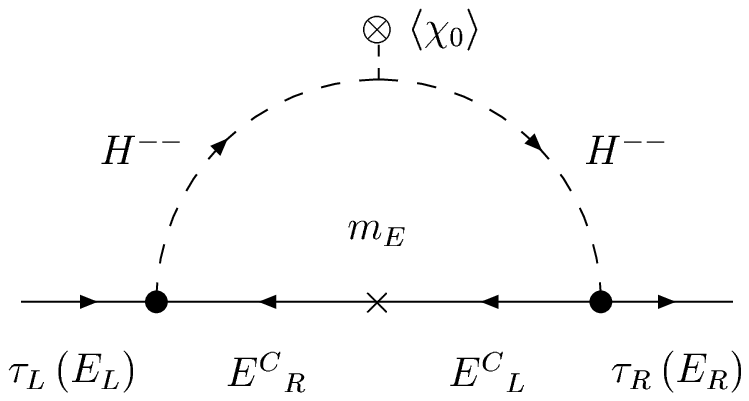}
\caption{\label{fig1} One loop diagram which induces the muon mass. There is a similar diagram where the heavy lepton $E$, in the internal lines, is replaced by the tau lepton.}
\end{figure}
\begin{subequations}
\begin{eqnarray}
{\cal U}^{\left(\ell\right)}_L & \approx & \left(\begin{array}{cccc}
-1/\sqrt{6} & 1/\sqrt{2} & 1/\sqrt{3} & 0 \\
\sqrt{2/3} & 0 & 1/\sqrt{3} & 0 \\
-1/\sqrt{6} & -1/\sqrt{2} & 1/\sqrt{3} & 0 \\
0 & 0 & 0 & -1
\end{array}\right),
\label{lmis1} \\
{\cal U}^{\left(\ell\right)}_R & \approx & \left(\begin{array}{cccc}
-1/\sqrt{6} & 1/\sqrt{2} & 1/\sqrt{3} & 0 \\
-1/\sqrt{6} & -1/\sqrt{2} & 1/\sqrt{3} & 0 \\
\sqrt{2/3} & 0 & 1/\sqrt{3} & 0 \\
0 & 0 & 0 & 1 
\end{array}\right)
\label{lmis}\end{eqnarray}\label{lm}\end{subequations}
calculated before the introduction of the corrections for electron and muon masses [see mass matrix (\ref{mat1})]. The matrices (\ref{lm}) are almost orthogonal and unitary. 

\section{Neutrino masses and mixings}
\label{sec4}

The scheme for neutrino mass and mixing generation is similar to the charged lepton one. The tau neutrino makes the role of the heavy charged lepton $E$ in the charged lepton sector. A difference relative to the charged sector is that in neutrinos sector a vanishing electron neutrino mass is still compatible with the present experimental data. Therefore, we go to keep $m_{\nu_e} = 0$. If it was necessary we could get a non zero mass for the electron neutrino through an explicit breaking of U(1)$_\Omega$, as we made for the electron. The mass matrix coming from the Lagrangian (\ref{lnu}) at the tree level is
\begin{figure}[ht]
\includegraphics{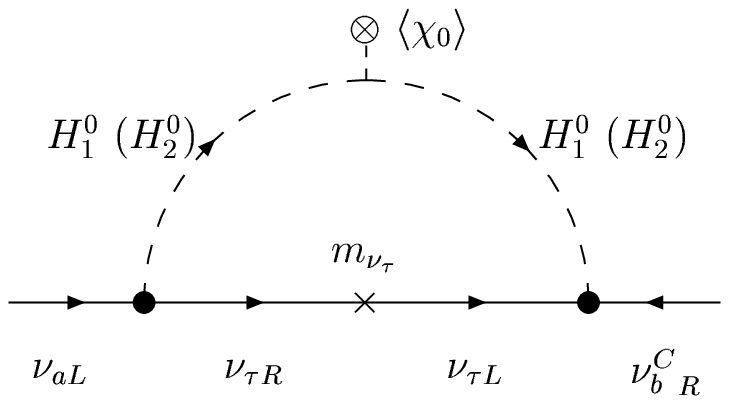}
\caption{\label{fig2} One loop diagram which induces the muon neutrino mass. As in the text, $a$ $=$ $e$, $\mu$, $\tau$.}
\end{figure}
\begin{equation}
{\cal M}_0^{(\nu)} = \left(\begin{array}{cccc} 0 & 0 & 0 & h_ev \\ 0 & 0 & 0 & h_\mu v \\ 0 & 0 & 0 & h_\tau v \\ h_ev & h_\mu v & h_\tau v & m \end{array}\right), \label{matnu0}
\end{equation}\noindent
in the $\left(\nu_e^\prime, \nu_\mu^\prime, \nu_\tau^\prime, N^{\prime C}\right)_L$ and $\left(\nu_e^{\prime C}, \nu_\mu^{\prime C}, \nu_\tau^{\prime C}, N^{\prime}\right)_R$ basis. Similarly as in the charged lepton sector, the neutrino mass matrix (\ref{matnu0}) gives the seesaw relation
\begin{equation}
m_Nm_{\nu_\tau} = -\left({\cal H}v\right)^2,
\label{massN}\end{equation}
where ${\cal H}^2 = h_e^2 + h_\mu^2 + h_\tau^2$ and the tau neutrino mass is given by
\begin{equation}
4m_{\nu_\tau} = m - \sqrt{m^2 + 4v^2{\cal H}^2}.
\label{ssnu}\end{equation}
The masses of the muon neutrino and of the electron neutrino vanish at the tree level, but the muon neutrino one rises as a radioactive correction at one loop level. \par
The transformations that diagonalize the neutrino mass matrix (\ref{matnu0}) are
\begin{widetext}
\begin{subequations}
\begin{eqnarray}
\nu^\prime_{eL} & = & \frac{1}{h}\left(\frac{h_eh_\tau}{\cal H}\nu_{eL} + h_\mu\nu_{\mu L}\right) - \frac{h_e}{\cal H}\left(-s_{\theta_\nu}\nu_{\tau L} + c_{\theta_\nu}{N^C}_L\right), \\ 
\nu^\prime_{\mu L} & = & \frac{1}{h}\left(\frac{h_\mu h_\tau}{\cal H}\nu_{eL} + h_e\nu_{\mu L}\right) + \frac{h_\mu}{\cal H}\left(-s_{\theta_\nu}\nu_{\tau L} + c_{\theta_\nu}{N^C}_L\right), \\ 
\nu^\prime_{\tau L} & = & \frac{1}{\cal H}\left[-h\nu_{eL} + h_\tau\left(-s_{\theta_\nu}\nu_{\tau L} + c_{\theta_\nu}{N^C}_L\right)\right], \\ {N^{\prime C}}_L & = & c_{\theta_\nu}\nu_{\tau L} + s_{\theta_\nu}{N^C}_L,
\end{eqnarray}
\label{eignul}
\end{subequations}
\end{widetext}
in the left-handed sector, where we are defining the parameter $h^2 = h_e^2 + h_\mu^2$. In the Eqs. (\ref{eignul}) we have
\begin{equation}
\tan{\left(\theta_\nu\right)} = \frac{-m + \sqrt{m^2 + 4v^2{\cal H}^2}}{2v{\cal H}}.
\end{equation}
In the right-handed sector the eigenstates ${\nu_e^{\prime C}}_R$, ${\nu_\mu^{\prime C}}_R$, ${\nu_\tau^{\prime C}}_R$ and $N^{\prime}_R$ transform as in Eqs. (\ref{eignul}), since that in the neutrino sector the mass matrix (\ref{matnu0}) is diagonalized by a unitary transformation. From the eigenstates (\ref{eigh}) and (\ref{eignul}) and the Lagrangian (\ref{lnu}) the relevant piece of the first order Lagrangian for neutrino mass terms is
\begin{widetext}
\begin{eqnarray}
{\cal L}_\nu^{\left(1\right)} & = & \frac{1}{v_W}\left\{\frac{1}{h}\left[\frac{\left(h_e + h_\mu\right)h_\tau - h^2}{\cal H}\overline{\nu_{eL}} + \left(h_e - h_\mu\right)\overline{\nu_{\mu L}}\right] - \frac{s_{\theta_\nu}}{\cal H}\left(h_e + h_\mu + h_\tau\right)\overline{\nu_{\tau L}}\right\}\times \cr
&& \times\left(c_{\theta_\nu}{\nu^C_\tau}_R + s_{\theta_\nu}N_R\right)\left(vH_1^0 + uH_2^0\right) + {\mbox{H. c.}}
\label{l1nu}\end{eqnarray}
\end{widetext}
The Lagrangian (\ref{l1nu}) leads to diagrams as in Fig. 2 that induces the finite neutrino mass terms
\begin{widetext}
\begin{eqnarray}
m_{ab}^{\left(\nu\right)} & = & -i\frac{\Theta w}{8\pi^2}m_{\nu_\tau}\sum_{\substack{i, j = 1, 2 \\ a, b = e, \mu, \tau}}X^{\left(j\right)*}_{\tau b}X^{\left(i\right)}_{a\tau} \times \cr
&& \times \frac{m_i^2m_{\nu_\tau}^2\ln{\left(m_i/m_{\nu_\tau}\right)} + m_j^2m_{\nu_\tau}^2\ln{\left(m_{\nu_\tau}/m_j\right)} + m_i^2m_j^2\ln{\left(m_j/m_i\right)}}{\left(m_i^2 - m_j^2\right)\left(m_{\nu_\tau}^2 - m_j^2\right)\left(m_{\nu_\tau}^2 - m_i^2\right)}.
\label{mN}\end{eqnarray}
\end{widetext}
In Eq. (\ref{mN}), $m_i$ are the Higgs boson masses and $\Theta$ is the strength of the trilinear interaction (see Fig. 2). $X^{\left(i\right)}$ are coefficients of the vertices in Fig. 2 and they can be read from the Lagrangian (\ref{l1nu}). \par
The values that we have chosen for the parameters were $h_e = h_\mu = h_\tau = -4i \times 10^{-13}$, and $m_\tau = 3 \times 10^{-2}$ eV, implying $m_N = 6.46 \times 10^{-3}$ eV [see Eqs. (\ref{massN}) and (\ref{ssnu})]. For simplicity we take $m_1 = m_2 = 200$ GeV. In this case, we have $\Theta = \Lambda vu/v_W^2$. The values for $u$, $w$ and $\Lambda$ come from of the numerical example of the charged leptons in Sec. \ref{sec3}. Thus, the neutrino mass matrix corrected up to the first order of the perturbative expansion is
\begin{widetext}
\begin{equation} {\cal M}_1^{(\nu)} \approx
\left(\begin{array}{cccc}
0 & 0 & 0 & 1.11 \times 10^{-2}\\
0 & 0 & 0 & 1.11 \times 10^{-2}\\
0 & 0 & -2.46i \times 10^{-8} & 1.11 \times 10^{-2} \\
1.11 \times 10^{-2} & 1.11 \times 10^{-2} & 1.11 \times 10^{-2} & 5.58 \times 10^{-2}
\end{array}\right),
\label{matnu1}\end{equation}
\end{widetext}
where we are neglecting corrections to zeroth order terms of the last column and the last line. The muon neutrino masses coming from the matrix (\ref{matnu1}) is $m_{\nu_\mu} = 1.21 \times 10^{-5}$ eV, while the first order tau neutrino and $m_N$ masses are not appreciably modified. We checked that these values are in agreement with the experimental results about neutrino oscillation, {\it i. e.}, $m_{\nu_e} \leq 3$ eV, $m_{\nu_\mu} \leq 1.9 \times 10^5$ eV, $m_{\nu_\tau} \leq 1.82 \times 10^7$ eV with $m_{\nu_\tau}^2 - m_N^2 \sim 3 \times 10^{-3}$ eV$^2$ from atmospheric and $m_{\nu_e}^2 - m_{\nu_\mu}^2 = \left(10^{-5} - 10^{-4}\right)$ eV$^2$ from solar neutrino oscillation data \cite{Hea02}. This numerical example leads to the neutrino mixing matrices, before the correction for the muon neutrino mass, as
\begin{eqnarray}
{\cal U}^{\left(\nu\right)} & \approx &
\left(\begin{array}{cccc}
1/\sqrt{6} & -1/\sqrt{2} & -1/\sqrt{3} & 0\\
-\sqrt{2/3} & 0 & -1/\sqrt{3} & 0\\
1/\sqrt{6} & 1/\sqrt{2} & -1/\sqrt{3} & 0\\
0 & 0 & 0 & 1 \end{array}\right).
\label{numis2}\end{eqnarray}
However, since the heavy singlet states do not couple to the standard $W^\pm$ bosons, we construct the mixing neutrino matrix by the product of the Hermitian conjugate of the $3 \times 4$ submatrix of ${\cal U}^{\left(\ell\right)}_L$ in (\ref{lmis1}) by the $3 \times 4$ submatrix of ${\cal U}^{\left(\nu\right)}$ in (\ref{numis2}) \cite{TD04}, {\it i. e.},
\begin{equation}
{\cal V}^{\left(\nu\right)} \approx \left(\begin{array}{cccc}
-1/2 & -0.87 & -10^{-4} & -2 \times 10^{-8} \\
-0.87 & -1/2 & 0 & 0 \\
2\times 10^{-18} & 0 & -6 \times 10^{-15} & 10^{-18} \\
10^{-4} & 0 & -1 & -2\times 10^{-4}
\end{array}\right)
\label{misnu}\end{equation}
A comparison of the mixing matrix (\ref{misnu}) with the Maki-Nakagawa-Sakata matrix for mixing of three standard neutrinos is not direct since here we have a fourth sterile neutrino involved in the mixing. However, the $3 \times 3$ submatrix corresponding to three standard neutrino mixings [obtained from the matrix (\ref{misnu}) by eliminating the last line and the last column] is of the LMA type, as is suggested by the present stage of the neutrino oscillation search \cite{Hea02}. 

\section{Comments and Conclusions}
\label{sec5}

We presented a scheme for generation of lepton mass based in a 3-3-1 model version which contains one charged heavy lepton singlet and one heavy neutrino singlet. The model obeys approximately an extra U(1)$_\Omega$ symmetry which is not spontaneously broken, but is slight violated by an explicit electron mass term. The heavy lepton masses satisfy seesaw relations, while the light masses, except the electron and the electron neutrino ones are generated by one loop radiative corrections scheme. We notice that in the realistic range of the parameters the muon mass expressed in the formula (\ref{massmu}) gives the relation $m_\mu m_\tau m_E \approx \left(f_ef^\prime_e + f_\mu f_\mu^\prime\right)uw\xi$. \par
In the context of the three neutrino states, data on oscillations and masses establish the particular quasi-bimaximal form to the neutrino mixing matrix \cite{Hea02}. A careful analysis of the experimental data furnishes the present range of the lepton mixing matrix as \cite{GG03}
\begin{equation}
\vert V_{\rm ex}\vert = \left(\begin{array}{ccc}
0.72 - 0.88 & 0.46 - 0.68 & < 0.22\\
0.25 - 0.65 & 0.27 - 0.73 & 0.55 - 0.84 \\
0.10 - 0.57 & 0.41 - 0.80 & 0.52 - 0.83
\end{array}\right).
\label{Vex}\end{equation}
When we compare the matrix (\ref{Vex}) with the 3$\times$3 mixing matrix obtained of Eq. (\ref{misnu}) by elimination of the fourth line and fourth column we can see that our numerical result for the lepton mixing is not in the experimental range. However, in this model the mixing matrix deserves a most detailed study. The numerical example that we take here is not only the possible one. Another fact that we must lead in account is that the mass matrix (\ref{mat0}) of the charged leptons is diagonalized by a biunitary transformation. The diagonalization process does not fix all the elements of the mixing matrices (\ref{lmis1}) and (\ref{lmis}). Three of them, it does not import which of them, remain free. Here, for sake of simplicity, we chosen ${\cal U}^{\left(\ell\right)}_L\left(3,2\right) = {\cal U}^{\left(\ell\right)}_L\left(4,4\right) = {\cal U}^{\left(\ell\right)}_R\left(4,4\right) = 0$. However, we have checked that the ${\cal U}^{\left(\ell\right)}_L\left(3,2\right)$, for example, can be used for improve the fit of our results. Moreover, we must to remember that the mixing matrices (\ref{lmis1}) and (\ref{numis2}) that we use to construct the matrix (\ref{misnu}) are approached, that is, they do not take in account the radiative corrections that generate the masses of muon and of the neutrino of muon.\par
Attempts to explain the shift from the exact bimaximal form for $\nu_e$, $\nu_\mu$ and $\nu_\tau$ are based on three effects: (i) from neutrino mixing becoming the charged lepton mixing \cite{RO04}, (ii) from radiative corrections \cite{MS03} and (iii) from the symmetry breaking \cite{GM02}. In our model the potential source of shift is the radiative correction processes leading to the muon and muon neutrino masses. The amount of the mixing in each eigenstate is function of the values of the parameters chosen. The electron mass, which induces a tiny breaking of the extra symmetry (\ref{carga}), contributes also to this shift, but with a smaller quantity. However, in our numerical example the LMA form is preserved in the sector of the three standard neutrinos, as is required by the experiments. In the charged sector, the mixing angles are also large [see Eqs. (\ref{lm})], so the concept of family in the lepton sector, in this sense, meaningless. \par
Our model is able to explain all the present experimental data on charged and neutral leptons, explaining the phenomenology within a reasonable scheme. This model naturally introduce de seesaw type relations, through an U(1) symmetry combining the good features of this mechanism with a radiative corrections scheme. The new mass scale is in the order of a few TeV. Therefore, this model can be confirmed or ruled out by the next generation of accelerators, what is a common feature of the class of 3-3-1 models.

\acknowledgements

One of us (M. D. T) thanks D. Fregolente for its inestimable cooperation in the beginning of this work and O. A. T. Dias, V. Pleitez and J. C. Montero for illuminating discussions.

\end{document}